\documentclass[twocolumn, showpacs, preprintnumbers, superscriptaddress, aps, prl, floatfix, tightenlines]{revtex4-1}         % TOUR TWO
\pdfoutput=1
\usepackage{hyperref}
\usepackage{amsmath}
\usepackage{cleveref}
\usepackage{CJKutf8}
\usepackage{latexsym}
\usepackage{amstext}
\usepackage{amsfonts}
\usepackage{dsfont}
\usepackage{color}
\usepackage{amssymb}
\usepackage{relsize}
\usepackage{amsxtra}
\usepackage{verbatim}
\usepackage{hyperref}
\usepackage{graphics}
\usepackage{graphicx}
\usepackage{multirow}

\AtBeginDvi{\input{zhwinfonts}}
\renewcommand{\thetable}{\arabic{table}}

\begin{document}
\begin{CJK*}{UTF8}{shiai}

\title{Enhanced magnetocaloric effect and magnetic phase diagrams of \\single-crystal GdCrO$_3$}

\author{Yinghao Zhu} %()}
%\email{yb87818@connect.um.edu.mo}
\affiliation{Joint Key Laboratory of the Ministry of Education, Institute of Applied Physics and Materials Engineering, University of Macau, Avenida da Universidade, Taipa, Macao SAR 999078, China}
\author{Pengfei Zhou} %()}
\affiliation{Joint Key Laboratory of the Ministry of Education, Institute of Applied Physics and Materials Engineering, University of Macau, Avenida da Universidade, Taipa, Macao SAR 999078, China}
\author{Tao Li} %()}
\affiliation{Neutron Scattering Technical Engineering Research Center, School of Mechanical Engineering, Dongguan University of Technology, Dongguan 523808, China}
\author{Junchao Xia} %()}
\affiliation{Joint Key Laboratory of the Ministry of Education, Institute of Applied Physics and Materials Engineering, University of Macau, Avenida da Universidade, Taipa, Macao SAR 999078, China}
\author{Si Wu} %()}
\affiliation{Joint Key Laboratory of the Ministry of Education, Institute of Applied Physics and Materials Engineering, University of Macau, Avenida da Universidade, Taipa, Macao SAR 999078, China}
\author{Ying Fu} %()}
\affiliation{Joint Key Laboratory of the Ministry of Education, Institute of Applied Physics and Materials Engineering, University of Macau, Avenida da Universidade, Taipa, Macao SAR 999078, China}
\author{Kaitong Sun} %()}
\affiliation{Joint Key Laboratory of the Ministry of Education, Institute of Applied Physics and Materials Engineering, University of Macau, Avenida da Universidade, Taipa, Macao SAR 999078, China}
\author{Qian Zhao} %()}
\affiliation{Joint Key Laboratory of the Ministry of Education, Institute of Applied Physics and Materials Engineering, University of Macau, Avenida da Universidade, Taipa, Macao SAR 999078, China}
\author{Zhen Li}
%\email{ailz268@126.com (Z. Li)}
\affiliation{Guangdong Provincial Engineering Research Center of Crystal and Laser Technology, Guangzhou, Guangdong 510632, China}
\affiliation{Department of Optoelectronic Engineering, Jinan University, Guangzhou, Guangdong 510632, China}
\author{Zikang Tang} %()}
\affiliation{Joint Key Laboratory of the Ministry of Education, Institute of Applied Physics and Materials Engineering, University of Macau, Avenida da Universidade, Taipa, Macao SAR 999078, China}
\author{Yinguo Xiao}
\email{y.xiao@pku.edu.cn (Y. Xiao)}
\affiliation{School of Advanced Materials, Peking University, Shenzhen Graduate School, Shenzhen 518055, China}
\author{Zhenqiang Chen}
\email{tzqchen@jnu.edu.cn (Z.-Q. Chen)}
\affiliation{Guangdong Provincial Engineering Research Center of Crystal and Laser Technology, Guangzhou, Guangdong 510632, China}
\affiliation{Department of Optoelectronic Engineering, Jinan University, Guangzhou, Guangdong 510632, China}
\author{Hai-Feng Li} %()}
\email{haifengli@um.edu.mo (H.-F. Li)}
\affiliation{Joint Key Laboratory of the Ministry of Education, Institute of Applied Physics and Materials Engineering, University of Macau, Avenida da Universidade, Taipa, Macao SAR 999078, China}

\date{\today}
\begin{abstract}

The crystalline structure, magnetism, and magnetocaloric effect of a GdCrO$_3$ single crystal grown with the laser-diode-heated floating-zone technique have been studied. The GdCrO$_3$ single crystal crystallizes into an orthorhombic structure with the space group $Pmnb$ at room temperature. Upon cooling, under a magnetic field of 0.1 T, it undergoes a magnetic phase transition at $T_{\textrm{N-Cr}} =$ 169.28(2) K with Cr$^{3+}$ ions forming a canted antiferromagnetic (AFM) structure, accompanied by a weak ferromagnetism. Subsequently, a spin reorientation takes place at $T_{\textrm{SR}} =$ 5.18(2) K due to Gd$^{3+}$-Cr$^{3+}$ magnetic couplings. Finally, the long-range AFM order of Gd$^{3+}$ ions establishes at $T_{\textrm{N-Gd}} =$ 2.10(2) K. Taking into account the temperature-(in)dependent components of Cr$^{3+}$ moments, we obtained an ideal model for describing the paramagnetic behavior of Gd$^{3+}$ ions within 30--140 K. We observed a magnetic reversal (positive $\rightarrow$ negative $\rightarrow$ positive) at 50 Oe with a minimum centering around 162 K. In the studied temperature range of 1.8--300 K, there exists a strong competition between magnetic susceptibilities of Gd$^{3+}$ and Cr$^{3+}$ ions, leading to puzzling magnetic phenomena. We have built the magnetic-field-dependent phase diagrams of $T_{\textrm{N-Gd}}$, $T_{\textrm{SR}}$, and $T_{\textrm{N-Cr}}$, shedding light on the nature of the intriguing magnetism. Moreover, we calculated the magnetic entropy change and obtained a maximum value at 6 K and ${\Delta}{\mu}_0H$ = 14 T, i.e., --${\Delta}S_{\textrm{M}} \approx$ 57.5 J/kg K. Among all \emph{R}CrO$_3$ (\emph{R} = $4f^n$ rare earths, $n =$ 7--14) compounds, the single-crystal GdCrO$_3$ compound exhibits the highest magnetic entropy change, as well as an enhanced adiabatic temperature, creating a prominent magnetocaloric effect for potential application in magnetic refrigeration.

\end{abstract}

\maketitle
\end{CJK*}

%\begin{widetext}

\section{I. INTRODUCTION}

The GdCrO$_3$ compound was initially synthesized in 1956 \cite{Geller1956}. Its crystalline structure was proved to be orthorhombic with the space group $Pbnm$ and lattice constants \emph{a} = 5.312 {\AA}, \emph{b} = 5.514 {\AA}, and \emph{c} = 7.611 {\AA} \cite{Geller1957}. Recently, the space group was determined to be $Pna2_1$ \cite{Mahana2018-2}. Later, its infrared and electronic absorption spectra were studied \cite{Rao1970}. Although the GdCrO$_3$ single crystal grown with a flux method may contain $\sim$1\% impurity, it was proposed that below $\sim$7 K, the ionic Cr$^{3+}$ magnetic sublattice underwent a spin reorientation from the $\Gamma_4$ (\emph{G}$_x$, \emph{A}$_y$, \emph{F}$_z$) to the $\Gamma_2$ (\emph{F}$_x$, \emph{C}$_y$, \emph{G}$_z$) magnetic structure. This was driven by the formation of the Gd$^{3+}$ magnetic sublattice and the Gd$^{3+}$-Cr$^{3+}$ couplings \cite{Wanklyn1969, Cooke1974}. The behavior of negative magnetization with a minimum centered around 25 K was observed in a polycrystalline GdCrO$_3$ sample, which was ascribed to the interactions between paramagnetic (PM) Gd$^{3+}$ moments and the canted Cr$^{3+}$ antiferromagnetic (AFM) moments \cite{Yoshii2001}. It was reported that the GdCrO$_3$ compound displayed an electric polarization of 0.7 $\mu$C/m$^2$ at \emph{E} = 2.25 kV/cm, appearing simultaneously with the formation of the Cr$^{3+}$ magnetic structure below $T_\textrm{N-Cr}$ \cite{Rajeswaran2012}. The distortion in the GdCrO$_3$ structure was proposed to be associated with the off-center displacement of Gd atoms together with octahedral rotations via displacement of oxygen ions \cite{Mahana2017-2, Mahana2018-2}. Magnetic refrigeration has risen to become a new civilian refrigeration technology \cite{Chen2001, Sari2014}. Besides the ferroelectric properties, the GdCrO$_3$ compound has attracted much attention and been believed to be a promising magnetorefrigerator material due to its high effective magnetic entropy change \cite{Yin2015-2, Yin2016, Mahana2018-1, Dash2018}.

In this paper, we have grown a GdCrO$_3$ single crystal using laser diodes with the floating-zone (FZ) technique. We characterized the crystalline structure with a room-temperature x-ray powder diffraction (XRPD) study and the magnetic properties with a physical property measurement system (PPMS DynaCool instrument, Quantum Design). Based on our measurements, we studied the magnetocaloric effect (MCE) and found that the GdCrO$_3$ single crystal investigated in this study displayed the highest magnetic entropy change among all \emph{R}CrO$_3$ compounds (\emph{R} = $4f^n$ rare earths, $n =$ 7--14). Moreover, we have built the magnetic phase diagrams as a function of applied magnetic field around the magnetic phase transitions of Gd$^{3+}$ and Cr$^{3+}$ ions, shedding light on a further understanding of the nature of the intriguing magnetism.

\section{II. EXPERIMENT}

Polycrystalline samples of the GdCrO$_3$ compound were synthesized using the conventional solid-state reaction method \cite{Li2008, Li2007-1, Li2007-2}. Chemically stoichiometric raw materials of Gd$_2$O$_3$ (Alfa Aesar, 99.9{\%}) and Cr$_2$O$_3$ (Alfa Aesar, 99.6{\%}) were milled and mixed by a Vibratory Micro Mill (FRITSCH PULVERISETTE 0) with an addition of 5--10\% Cr$_2$O$_3$ compound. The mixture was calcined twice at ambient air pressure: One time was at 1100 $^{\circ}$C for 24 h, and the other was at 1200 $^{\circ}$C for 36 h. We grew GdCrO$_3$ single crystals with a laser diode FZ furnace (model: LD-FZ-5-200W-VPO-PC-UM) \cite{Zhu2019-1,Zhu2020,Wu2020-1}.

We pulverized a small piece of the single crystal to check the phase purity and determine the room-temperature crystalline structure with XRPD employing copper $K_{\alpha1}$ (1.54056 {\AA}) and $K_{\alpha2}$ (1.544390 {\AA}) with a ratio of 2{:}1 as the radiation. The XRPD pattern was collected at 2$\theta =$ 20--90$^{\circ}$ with a step size of 0.02$^{\circ}$. We used the software FULLPROF SUITE \cite{Carvajal1993} to refine the collected XRPD data. We modeled the Bragg peak shape with a pseudo-Voigt function and used a linear interpolation between automatically selected data points to estimate the background contribution. The refining parameters are scale factor, zero shift, background contribution, peak shape parameters, asymmetry, preferred orientation, lattice constants, and atomic positions.

The measurements of dc magnetization and specific heat were carried out on a PPMS DynaCool instrument using the vibrating sample magnetometry and the heat capacity options, respectively. The dc magnetization at an applied magnetic field of 500 Oe was measured with two modes at 1.8--300 K: One was after cooling with 0 Oe, i.e., zero-field cooling (ZFC), and the other was at $\mu_0H =$ 500 Oe, i.e., field cooling (FC). To clearly show the effect of applied magnetic field on magnetic structures of the two magnetic ions Gd$^{3+}$ and Cr$^{3+}$, ZFC magnetization measurements at different fields as a function of temperature in two ranges (1.8--8 and 165--172 K) were performed. ZFC magnetization measurements from --14 to 14 T at different temperatures were carried out either in a mode of the magnetic hysteresis loop or as a function of increasing magnetic field. In order to gain the adiabatic temperature change, the specific heats at 1.8--300 K were measured at magnetic fields of 0, 0.5, 1, 2, 3, 5, 6, 8, 10, 12, and 14 T.

\section{III. Results and discussion}

\subsection{A. Structural study}

\begin{figure*} [!t]
\centering
\includegraphics[width=0.88\textwidth] {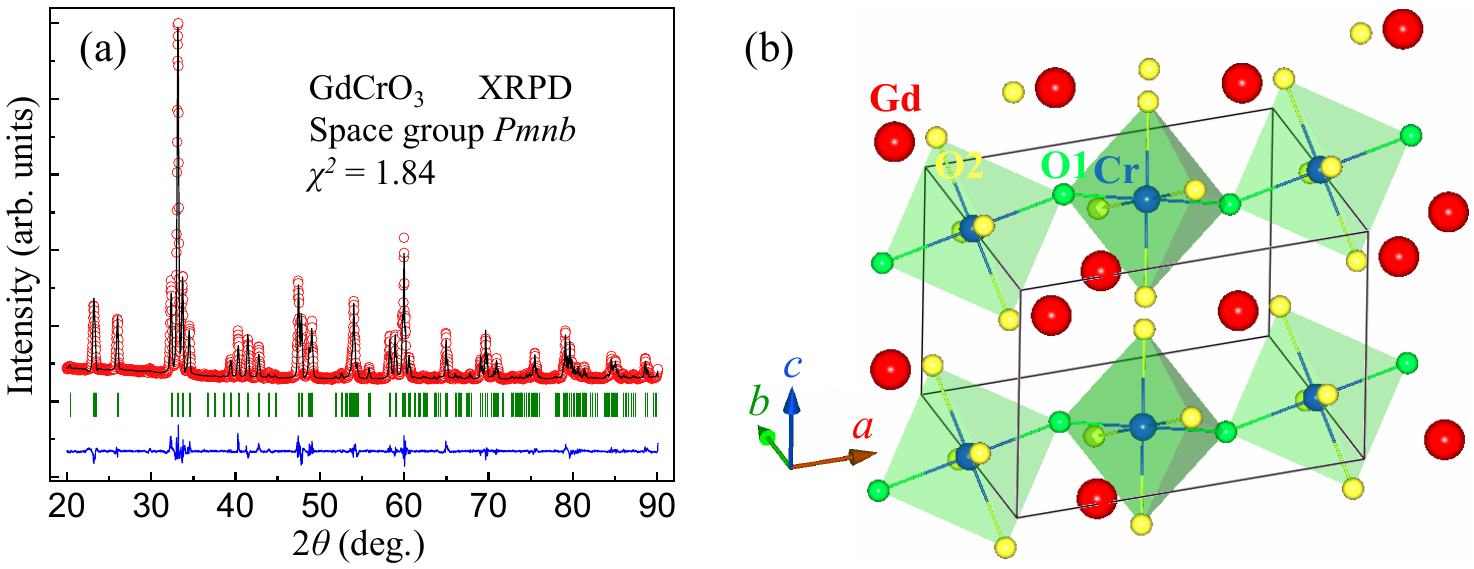}
\caption{(color online)
(a) Observed (circles) and refined (solid line) XRPD patterns collected with a pulverized GdCrO$_3$ single crystal at room temperature. Vertical bars mark the positions of Bragg peaks. The bottom curve represents the difference between observed and refined XRPD patterns.
(b) Crystal structure of the GdCrO$_3$ compound in one unit cell (solid lines) with space group \emph{Pmnb} (No. 62). The Gd, Cr, O1, and O2 ions are labeled.
}
\label{str}
\end{figure*}

To study the crystalline structure of our grown GdCrO$_3$ single crystal, we pulverized a small piece of the GdCrO$_3$ single crystal and carried out a XRPD experiment at room temperature. Figure~\ref{str}(a) shows the collected and refined patterns. Within the present experimental accuracy, the data can be well indexed with space group \emph{Pmnb}. The corresponding crystalline structure in one unit cell is exhibited in Fig.~\ref{str}(b), and the refined structural information is listed in Table~\ref{str-para}. The low values of the goodness of refinement validate our FULLPROF refinements. Our refined room-temperature lattice constants of the pulverized GdCrO$_3$ single crystal are \emph{a} = 7.6041(3) {\AA}, \emph{b} = 5.5255(2) {\AA}, and \emph{c} = 5.3102(2) {\AA}, consistent with previously reported values from a study with the polycrystalline GdCrO$_3$ compound \cite{Yoshii2001}.

\subsection{B. Magnetic phase transitions}

\begin{table*}[!t]
\renewcommand*{\thetable}{\Roman{table}}
\caption{Refined structural parameters, including lattice constants, unit-cell volume \emph{V}, atomic positions, and goodness of refinement, from room-temperature XRPD with a pulverized GdCrO$_3$ single crystal (orthorhombic, space group \emph{Pmnb}, No. 62, \emph{Z} = 4). The Wyckoff sites of all atoms are listed. We kept the atomic occupation factors (OCs) during FULLPROF refinements. The numbers in parentheses are the estimated standard deviations of the last significant digit. $R_\textrm{p}$ = 3.28, $R_\textrm{wp}$ = 4.64, $R_\textrm{exp}$ = 3.43, and $\chi^2$ = 1.84.}
\label{str-para}
\begin{ruledtabular}
\begin{tabular} {llllll}
$a$ (\AA)     & $b$ (\AA)         & $c$ (\AA)      & $V$ ({\AA}$^3$)    & \multicolumn{2}{l} {$\alpha (= \beta = \gamma)$ (deg)}                                          \\
\hline
7.6041(3)     & 5.5255(2)         & 5.3102(2)      & 223.12(2)          & \multicolumn{2}{l} {90}                                                                         \\
Atom          & Site              & \emph{x}       & \emph{y}           & \emph{z}            & OCs                                                                       \\
\hline
Gd            & 4\emph{c}         & 0.25           & 0.0588(2)          & 0.0151(3)           & 0.5                                                                       \\
Cr            & 4\emph{b}         & 0.00           & 0.00               & 0.50                & 0.5                                                                       \\
O1            & 4\emph{c}         & 0.25           & 0.4705(19)         & 0.1146(19)          & 0.5                                                                       \\
O2            & 8\emph{d}         & 0.0530(10)     & 0.2784(17)         & --0.2935(16)        & 1.0                                                                       \\
\end{tabular}
\end{ruledtabular}
\end{table*}

\begin{figure*} [!t]
\centering
\includegraphics[width = 0.88\textwidth] {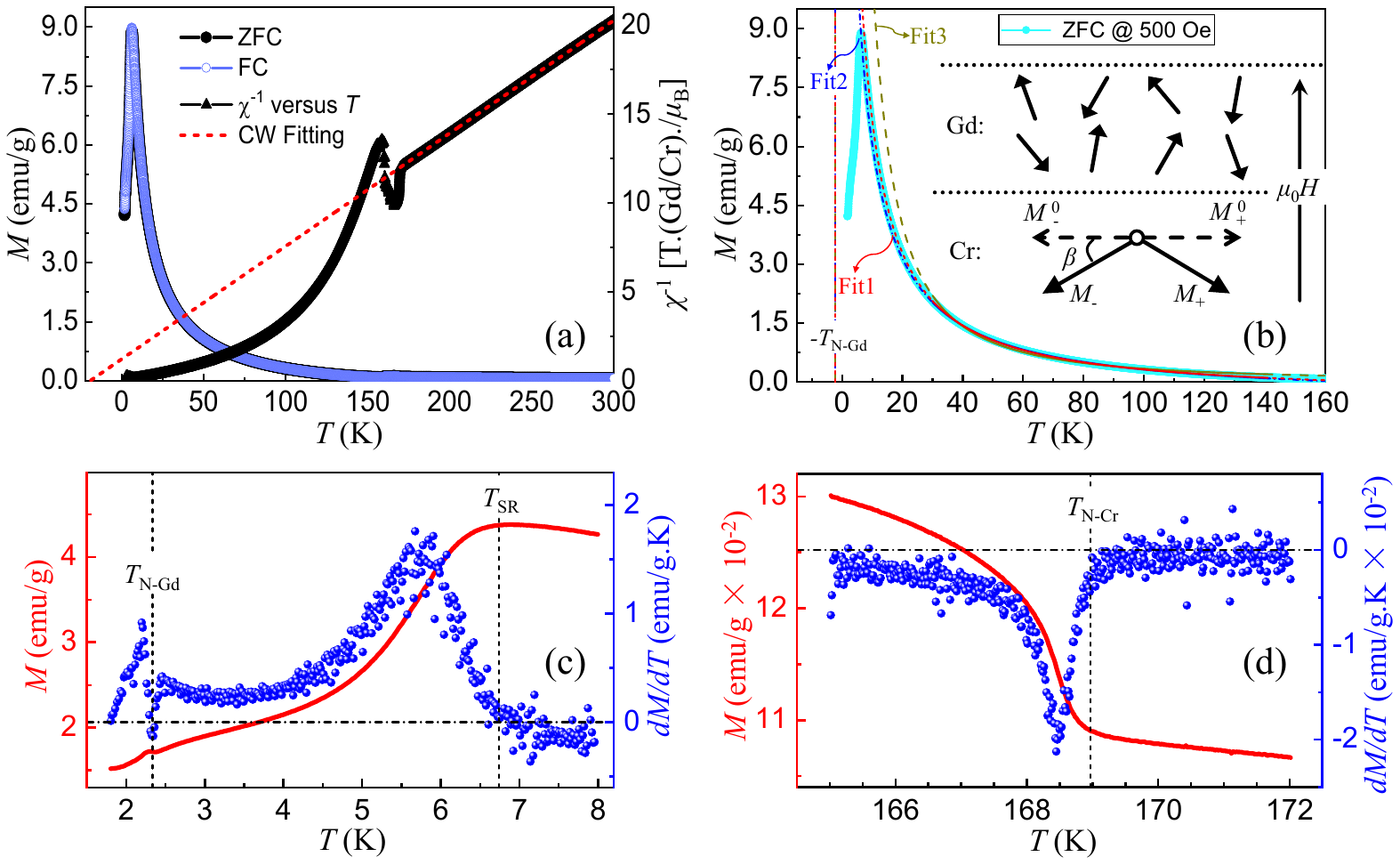}
\caption{(color online)
Representative magnetization measurements of a small piece of GdCrO$_3$ single crystal with random crystallographic orientations.
(a) ZFC (solid circles) and FC (void circles) magnetization \emph{M} (left axis) and the corresponding ZFC inverse magnetic susceptibility ${\chi}^{-1}$ (solid triangles; right axis) at an applied magnetic field of 500 Oe as a function of temperature in the range of 1.8--300 K. The dashed line represents the fit with a CW law.
(b) ZFC \emph{M} versus temperature (solid circles) measured at 500 Oe from 1.8 to 160 K. The solid lines denote fit 1 [with Eq.~(\ref{realMT})], fit 2 [with Eq.~(\ref{MT})], and fit 3 [with Eq.~(\ref{realMT})] in the temperature range of 30--140 K. They were extrapolated to the whole temperature regime [--$T_{\textrm{N-Gd}}$, 160 K] and are shown as the short-dashed line (fit1), dash-dotted line (fit 2), and long-dashed line (fit 3). The inset schematically shows spin configurations of Gd$^{3+}$ and Cr$^{3+}$ ions within 30--140 K. See details in the text.
(c) ZFC \emph{M} (left axis), as well as the corresponding $dM/dT$ (right axis), versus temperature in the range of 1.8--8 K at 200 Oe. $T_{\textrm{N-Gd}}$ points out the magnetic transition temperature of Gd$^{3+}$ ions, which we define as the temperature point where the slope of the $M-T$ curve is minimum. $T_{\textrm{SR}}$ indicates the spin reorientation (SR) temperature of Cr$^{3+}$ ions, which we define as the temperature point from which the slope of the $M-T$ curve changes from negative to positive upon cooling.
(d) ZFC \emph{M} (left axis) and its slope $dM/dT$ (right axis) versus temperature in the range of 165--172 K at an applied magnetic field of 600 Oe. $T_{\textrm{N-Cr}}$ implies the magnetic transition temperature of Cr$^{3+}$ ions, which we define as the temperature point at which a kink appears in the slope of the $M-T$ curve upon cooling.}
\label{magT}
\end{figure*}

Figure~\ref{magT} shows the measured magnetization as a function of temperature. As shown on the left axis of Fig.~\ref{magT}(a), as temperature decreases from 300 to 1.8 K, there is a smooth increase in the magnetization with an anomaly appearing around $T_\textrm{N-Cr}$ [Fig.~\ref{magT}(d)]. At $T_\textrm{N-Cr}$, we observed a small sharp increase [Fig.~\ref{magT}(d)]. This is by far clearer in the inverse magnetic susceptibility $\chi^{-1}$, as shown on the right axis of Fig.~\ref{magT}(a). The magnetization increases smoothly again until around 25 K. We observed a maximum at $T_\textrm{SR}$ [Fig.~\ref{magT}(c)]. Upon further cooling, there exists a kink at $T_\textrm{N-Gd}$ [Fig.~\ref{magT}(c)]. These anomalies are attributed to magnetic phase transitions. The first anomaly is related to Cr$^{3+}$ ions, and $T_\textrm{N-Cr} \approx$ 168.97 K at 0.06 T. The second one is ascribed to the spin reorientation of Cr$^{3+}$ ions due to the gradual formation of Gd$^{3+}$ moments \cite{Cooke1974}, and $T_\textrm{SR} \approx$ 6.74 K at 0.02 T. Taking into account the fact that the ordering of 4\emph{f} magnetic Gd$^{3+}$ ions requires much lower temperatures \cite{Jensen1991, Zhang2013}, the third one thus corresponds to the formation of a long-range-ordered Gd$^{3+}$ magnetic structure, and $T_\textrm{N-Gd} \approx$ 2.33 K at 0.02 T. The inverse magnetic susceptibility $\chi^{-1}$ in a pure PM state observes well with the Curie-Weiss (CW) law
\begin{eqnarray}
\chi^{-1}(T) = \frac{3k_B(T - \Theta_{\textrm{CW}})}{N_A \mu^2_{\textrm{eff}}},
\label{CWLaw}
\end{eqnarray}
where $k_B =$ 1.38062 $\times$ 10$^{-23}$ J/K is the Boltzmann constant, ${\Theta}_\textrm{CW}$ is the PM CW temperature, $N_A =$ 6.022 $\times$ 10$^{23}$ mol$^{-1}$ is Avogadro's constant, and $\mu_{\textrm{eff}}$ = $g\mu_\textrm{B} \sqrt{J(J + 1})$ is the effective PM moment. We fit the magnetization in the temperature range of 200--300 K ($> T_\textrm{N-Cr} > T_\textrm{N-Gd}$) to Eq.~(\ref{CWLaw}) and extrapolated the fit down to $M({\Theta}_\textrm{CW}) =$ 0, as shown on the right axis of Fig.~\ref{magT}(a). This results in an effective PM moment ${\mu}_{\textrm{eff}}$ = 8.40(9) ${\mu}_\textrm{B}$ and a PM CW temperature ${\Theta}_\textrm{CW}$ = --20.33(4) K. It is stressed that these values correlate with the PM behaviors of both Gd$^{3+}$ and Cr$^{3+}$ ions. Here, the extracted ${\mu}_{\textrm{eff}}$ = 8.40(9) ${\mu}_\textrm{B}$ is a little larger than the previously reported value of $\sim$8.2 ${\mu}_\textrm{B}$ from a study with the polycrystalline GdCrO$_3$ compound \cite{Sardar2011}, indicating a better quality of the single-crystal GdCrO$_3$ sample. For Gd$^{3+}$ ions (shell 4$f^7$, quantum numbers $S = \frac{7}{2}$, $L =$ 0, and $J = \frac{7}{2})$, the size of the theoretical (theo.) effective PM moment is 7.94 ${\mu}_\textrm{B}$, while for Cr$^{3+}$ ions, $\mu_{\textrm{eff{\_}theo.}} =$ 3.873 ${\mu}_\textrm{B}$ \cite{Zhu2020-2}, therefore, $\mu_{\textrm{eff{\_}theo.}} = \sqrt{7.94^2 + 3.873^2}$ ${\mu}_\textrm{B}$ = 8.834 ${\mu}_\textrm{B}$ for the GdCrO$_3$ compound. This theoretical value is $\sim$5.17\% larger than the corresponding experimental value of 8.40(9) ${\mu}_\textrm{B}$, which indicates that vacancies probably exist in the Gd and/or Cr atomic sites. We calculated schematically the magnetic frustration parameter $f = \frac{|\Theta_{\textrm{CW}}|}{T_\textrm{N}}$ \cite{Zhu2020}; for Cr$^{3+}$ ions at 0.05 T, $f_{\textrm{Cr}} \approx$ 0.12. Compared with the YCrO$_3$ compound within which the Y$^{3+}$ ions are nonmagnetic and $f_{\textrm{Cr}} \approx$ 3.06 \cite{Zhu2020-2}, the competing degree of AFM and ferromagnetic interactions in GdCrO$_3$ compound is much weaker. The introduction of magnetic Gd$^{3+}$ ions in the GdCrO$_3$ compound has a strong effect on the magnetic structure of Cr$^{3+}$ ions, in agreement with the appearance of $T_\textrm{SR}$.

\begin{table*}[!t]
\renewcommand*{\thetable}{\Roman{table}}
\caption{Fit values of the parameters $M_\textrm{BG}$ and $\gamma$ while modeling the temperature-dependent ZFC magnetization data of the GdCrO$_3$ single crystal (measured at 7--30 K and 500 Oe) with Eq.(~\ref{realMT}). We divided the whole temperature range into five regimes (see details in the text). During the refinements, we fixed $m =$ 114.17(51) emu K /g and $\Theta_{\textrm{CW}}$ = --2.33 K. The numbers in parentheses are the estimated standard deviations of the last significant digit.}
\label{gamma}
\begin{ruledtabular}
\begin{tabular} {lllllllllllllllllllll}
$T$ regime (K)                        &&&& 7--10        &&&& 10--15      &&&& 15--20         &&&& 20--25         &&&& 25--30            \\
\hline
$M_\textrm{BG}$ (emu/g)               &&&& 1.312(12)    &&&& 0.050(8)    &&&& --0.239(10)    &&&& --0.306(12)    &&&& --0.329(15)       \\
$\gamma$                              &&&& 1.229(1)     &&&& 1.144(1)    &&&& 1.122(1)       &&&& 1.115(1)       &&&& 1.112(2)          \\
\end{tabular}
\end{ruledtabular}
\end{table*}

\begin{figure*} [!t]
\centering
\includegraphics[width = 0.88\textwidth] {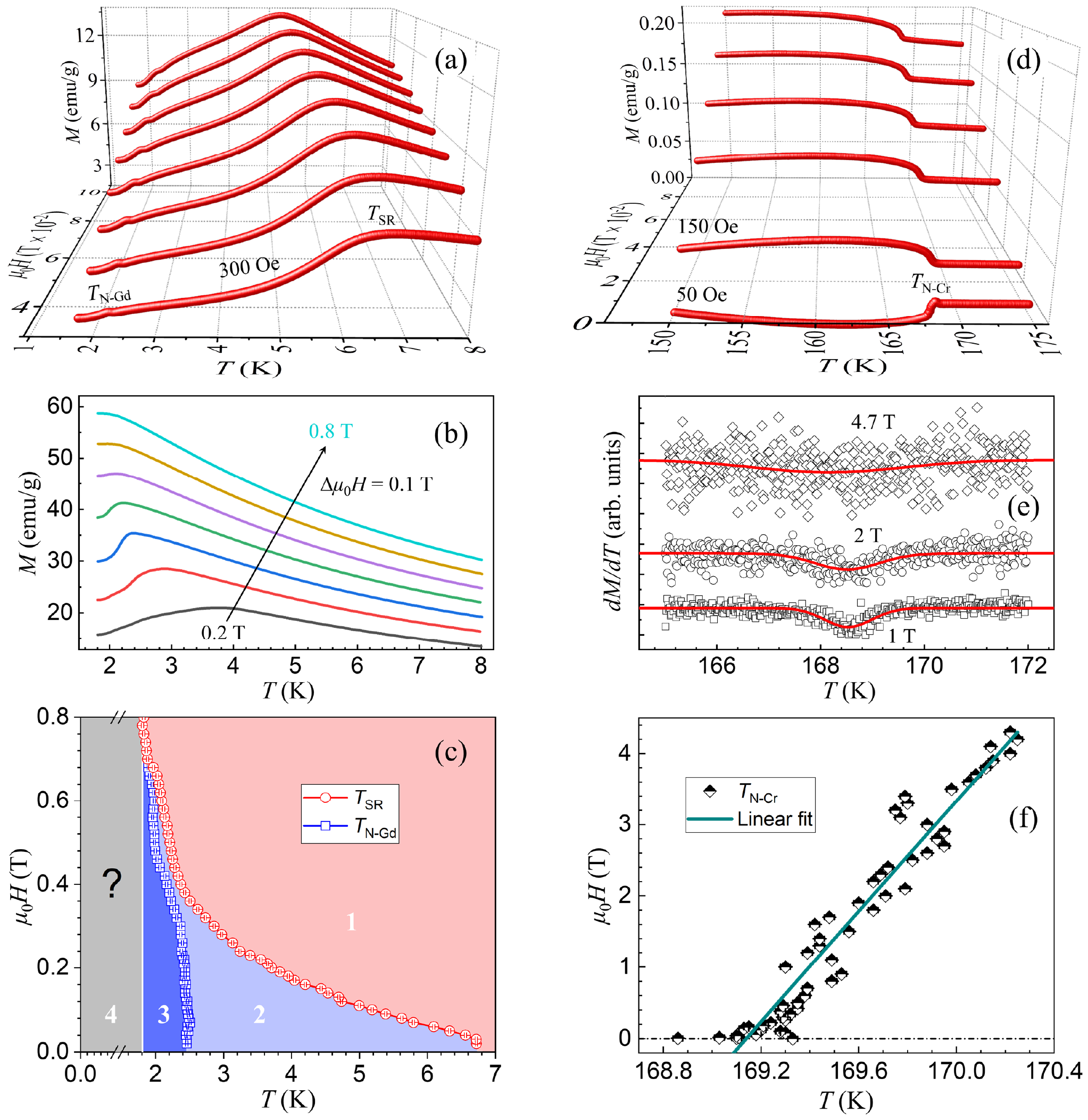}
\caption{(color online)
ZFC \emph{M} as a function of temperature from 1.8 to 8 K at applied magnetic fields of
(a) 300--1000 Oe with a step size of 100 Oe and (b) 0.2--0.8 T with $\Delta\mu_0H =$ 0.1 T.
(c) Applied magnetic-field- and temperature-dependent phase diagram of $T_{\textrm{N-Gd}}$ and $T_{\textrm{SR}}$.
(d) ZFC \emph{M} versus temperature from 150 to 175 K at applied magnetic fields of 50, 150, 400, 600, 800, and 1000 Oe.
(e) Slope $dM/dT$ (symbols) of the $M-T$ curve at 1, 2, and 4.7 T. The solid lines are fits with a modified Gaussian function, as guides to the eye.
(f) $T_{\textrm{N-Cr}}$ as a function of applied magnetic field (symbols). We fit tentatively the data with a linear function (solid line).}
\label{phasediag}
\end{figure*}

To analyze individually the PM behavior of Gd$^{3+}$ ions, we took the magnetization data in the temperature range of 30--140 K. This thermal range was within the interval ($T_\textrm{N-Gd}$, $T_\textrm{N-Cr}$) (Fig.~\ref{magT}), far above $T_\textrm{N-Gd}$ (to ensure that Gd$^{3+}$ ions were indeed in a PM state) and $\sim$20 K below $T_\textrm{N-Cr}$ (to weaken the effect of ordered Cr$^{3+}$ ions as much as possible). We first fit tentatively the data with
\begin{eqnarray}
\emph{M} = M_{\textrm{BG}} + \frac{m}{T - \Theta_{\textrm{CW}}},
\label{MT}
\end{eqnarray}
where $M_{\textrm{BG}}$ is the contribution from background (BG) magnetization that includes actual BG magnetization from the instrument and sample holder, as well as the glue, the temperature-independent diamagnetism components of the Gd$^{3+}$ and Cr$^{3+}$ ions, and the temperature-independent net magnetization of the Cr$^{3+}$ magnetic sublattice, and \emph{m} is a constant. Similar modeling strategies were used previously \cite{Cooke1974, Yoshii2001, Yin2015-2, Fita2019, Kumar2017}. The values of the diamagnetism of Gd$^{3+}$ and Cr$^{3+}$ ions are $\sim$--2.0 $\times$ 10$^{-5}$ and $\sim$--1.1 $\times$ 10$^{-5}$ emu/mol \cite{Bain2008}, respectively, which could be neglected reasonably. Since the Gd$^{3+}$ magnetic sublattice seems to form a long-range AFM order below $T_\textrm{N-Gd}$, most likely, the frustration parameter of Gd$^{3+}$ ions $f = \frac{|\Theta_{\textrm{CW}}|}{T_\textrm{N}^{\textrm{Gd}}} \approx$ 1, from which we deduced $\Theta_{\textrm{CW}}^{\textrm{Gd}} \approx -T_\textrm{N}^{\textrm{Gd}} \sim$--2.33 K. By forcing $\Theta_{\textrm{CW}}^{\textrm{Gd}} =$ --2.33 K, we refined the data with Eq.~(\ref{MT}) and obtained $M_{\textrm{BG}} =$ --0.473(33) emu/g and $m =$ 80.823(22) emu K/g, and the resultant fit is shown as fit2 in Fig.~\ref{magT}(b). It is worth noting that in Eq.~(\ref{MT}), the net magnetization of the Cr$^{3+}$ magnetic sublattice is supposed to be temperature independent, which is true only at low enough temperatures. For example, achieving this stage for the YCrO$_3$ single crystal, it requires temperatures at least below $\sim$50 K \cite{Zhu2020-2}.

\begin{figure*} [!t]
\centering
\includegraphics[width = 0.88\textwidth] {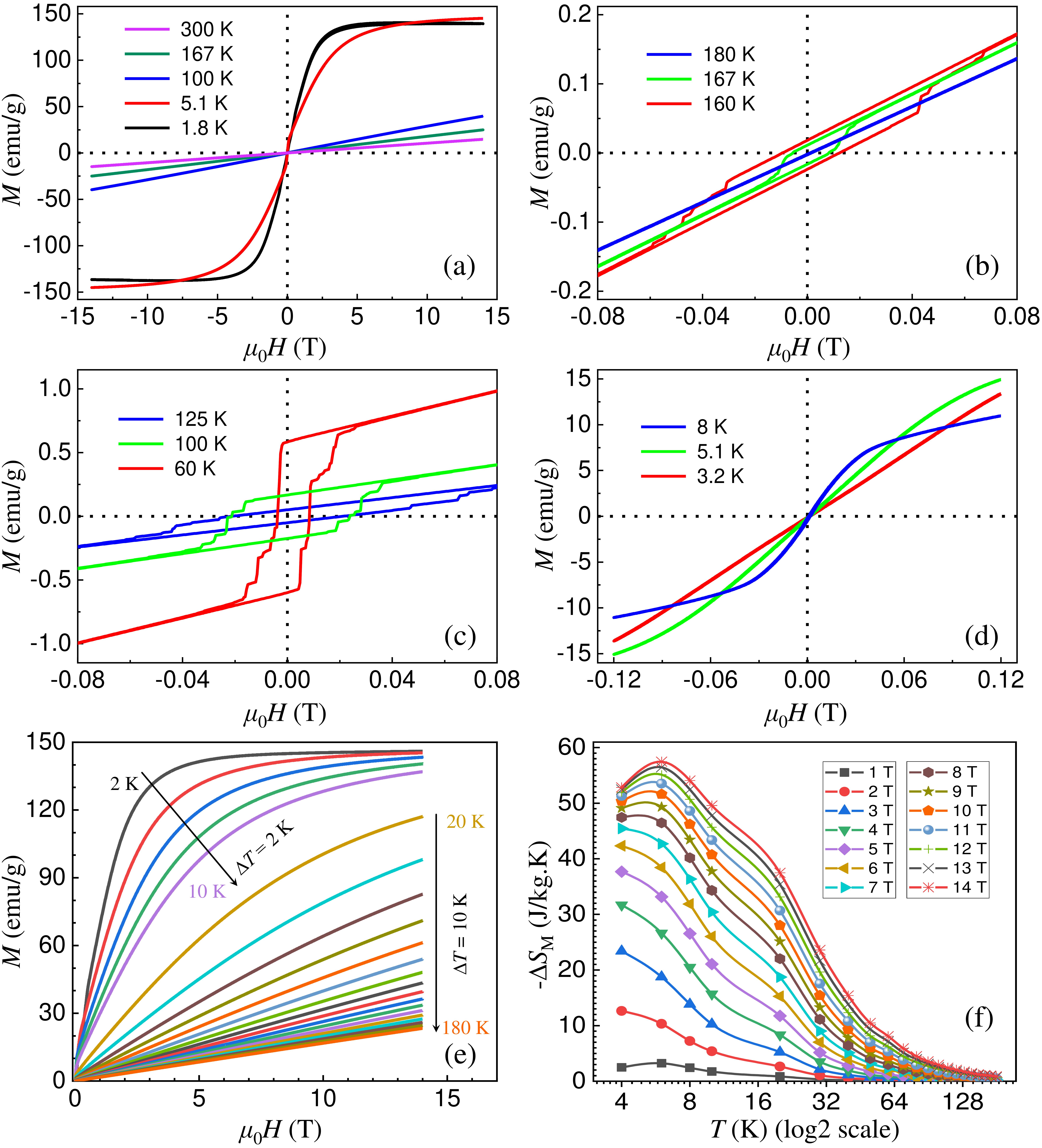}
\caption{(color online)
(a)-(d) Representative ZFC isothermal magnetization versus applied magnetic field.
(a) The field range is from --14 to 14 T, and the studied temperature points are at 1.8, 5.1, 100, 167, and 300 K.
The measured temperature points are
(b) 160, 167, and 180 K,
(c) 60, 100, and 125 K, and
(d) 3.2, 5.1, and 8 K. For (b--d), the magnetic fields are from --1.2 to 1.2 T.
(e) Representative ZFC magnetization as a function of applied magnetic field in the range of 0--14 T at temperatures of 2--10 K (step size of 2 K) and 10--180 K (step size of 10 K).
(f) Extracted magnetic entropy versus temperature in the thermal range of 4--180 K (with log$_\textrm{2}$ scale) at $\mu_0H =$ 1--14 T with an interval of 1 T. The solid lines are guides to the eye.}
\label{magH}
\end{figure*}

\begin{figure*} [!t]
\centering
\includegraphics[width = 0.88\textwidth] {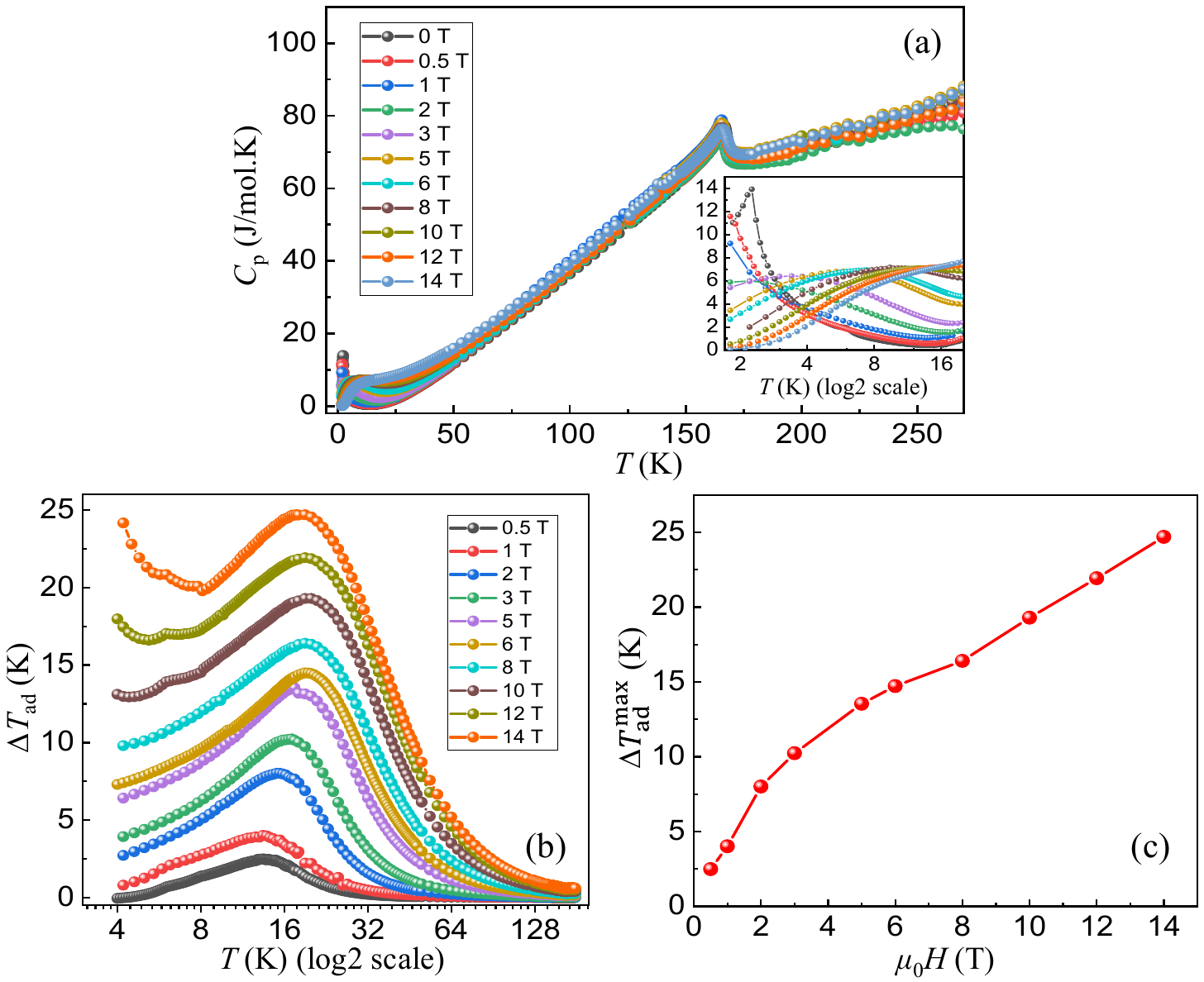}
\caption{(color online)
(a) Temperature-dependent specific heat measured at applied magnetic fields of 0--14 T. The inset exhibits specific heats within the temperature range of 1.8--20 K with a log$\textrm{2}$ scale.
(b) Temperature-dependent adiabatic temperature change $\Delta$$T_\textrm{ad}$ with applied magnetic fields from 0.5 to 14 T.
(c) Maximum of adiabatic temperature change $\Delta$$T_\textrm{ad}^\textrm{max}$ versus applied magnetic field.
In (a)--(c), the solid lines are guides to the eye.}
\label{Magcal}
\end{figure*}

For ferromagnets, it has been proved that the temperature-dependent magnetic susceptibility observes a power law $\chi_0^{-1}\propto(T-T_\textrm{C})^{\gamma}$ at temperatures slightly above the Curie temperature $T_\textrm{C}$, where the exponential parameter $\gamma$ usually acts as a criterion for distinguishing the Heisenberg system ($\gamma = \frac{4}{3}$) from the Ising one ($\gamma = \frac{5}{4}$) \cite{Domb1957, Domb1961}. When $\gamma$ = 1, it correlates with the CW model that is derived from the molecular-field theory; when $\gamma$ $>$ 1, it becomes a modified nonlinear formula that relates to the short-range spin correlations during the formation of the spin order \cite{Noakes1964, Kouvel1964}. For antiferromagnets, there also exists a power-law model in describing the staggered magnetic susceptibility $\chi$(\emph{Q}) $\varpropto\alpha$($T-T_\textrm{N}$)$^{-\gamma}$ below $T_\textrm{N}$ \cite{Moriya1962}; when $\alpha$ = 1, one may get the best fit \cite{Ishikawa1973}

Since $T_{\textrm{N-Gd}}$ (2.33 K) $\ll$ $T_{\textrm{N-Cr}}$ (168.97 K), when the magnetic structure of Cr$^{3+}$ ions is established below $T_{\textrm{N-Cr}}$, the staggered magnetization could induce short-range Gd$^{3+}$ magnetic orders due to possible Gd$^{3+}$-Cr$^{3+}$ couplings \cite{Dash2018, Jaiswal2010, Tripathi2019}. To model these complicated magnetic behaviors, we modified Eq.~(\ref{MT}) to
\begin{eqnarray}
\emph{M} = M_{\textrm{BG}} + \frac{m}{(T - \Theta_{\textrm{CW}})^\gamma},
\label{realMT}
\end{eqnarray}
where $\gamma$ is a parameter representing both the short-range correlations of Gd$^{3+}$ ions and the temperature-dependent component of the net magnetization of the Cr$^{3+}$ magnetic sublattice. This could be enhanced by the formation of magnetic polarons due to Gd$^{3+}$-Cr$^{3+}$ ionic couplings \cite{Li2012, Zhang2020}. We took the following strategies for refinements in the thermal range of 30--140 K: (i) First, we kept $M_{\textrm{BG}} =$ 0 and refined $m$ and $\gamma$, which resulted in $m =$ 618.1(12) emu K/g, and $\gamma =$ 1.6219(5), and the fitting result is shown as fit 3 in Fig.~\ref{magT}(b). (ii) While releasing $M_{\textrm{BG}}$, $m$, and $\gamma$, we refined them simultaneously, leading to $M_{\textrm{BG}} =$ --0.359(22) emu/g, $m =$ 114.17(51) emu.K./g, and $\gamma =$ 1.108(1), the best fit [fit 1 in Fig.~\ref{magT}(b)]. This best refinement is supported by the following facts: (i) $T =$ 30 K is a little above the temperature point at $\sim$24.51 K at which a kink appears in the slope of the $M-T$ curve and from where ZFC $M$ increases sharply upon cooling due to the formation of AFM Gd$^{3+}$ sublattice moments. (ii) While extrapolating these fits to the temperature range from --2.33 to 160 K [Fig.~\ref{magT}(b)], the calculated values of $M$($T$) [dash-dotted line (fit 2) and long-dashed line (fit 3)] deviate too much from the measured data below 30 K and above 140 K. It is noted that for the best fit, fit 1, the refined $M_{\textrm{BG}}$ is still negative, which is much smaller than the values of diamagnetism of Gd$^{3+}$ and Cr$^{3+}$ ions, and $\gamma >$ 1. As shown in the inset of Fig.~\ref{magT}(b), one possible configuration of spin moments for 30--140 K is as follows (i) The magnetic Gd$^{3+}$ ions stay in a PM state; that is, all spins are theoretically aligned randomly with potential short-range correlations induced possibly by the formation of the magnetic order of Cr$^{3+}$ ions. (ii) The formed Cr$^{3+}$ AFM sublattice moments ($M_+$ and $M_-$) are canted downward from their AFM axis $M^0_+M^0_-$ with an angle of $\beta$ so that the net Cr$^{3+}$ and Gd$^{3+}$ moments are in opposite directions, leading to a negative value of $M_{\textrm{BG}}$. It is pointed out that mere the applied magnetic field of 500 Oe itself is hard to make an AFM canting \cite{HFLi2016} and produce such a large net negative magnetization from the Cr$^{3+}$ magnetic sublattice, consistent with the formation of a canted AFM Cr$^{3+}$ structure.

\begin{table*}[!t]
\renewcommand*{\thetable}{\Roman{table}}
\caption{Comparison of the MCE in different \emph{R}CrO$_3$ compounds (\emph{R} = $4f^n$ rare earths, $n =$ 7--14). Here PC = polycrystal, SC = single crystal, FZM = floating-zone method, FLM = flux method, \emph{T} = temperature, Lit. = the literature, and TS = this study.}
\label{MCE}
\begin{ruledtabular}
\begin{tabular} {lllllllll}
Compound       & 4$f^n$   & Form         & \multicolumn{3}{c}{$-{\Delta}S_{\textrm{M}}$}                                             & ${\Delta}{\mu}_0H$     & \emph{T}      & Ref.                         \\
               &          &              & \multicolumn{3}{c}{(J/kg K)}                                                              & (T)                    & (K)           &                             \\
               &          &              & From Lits.                                         & From TS         & Improved           &                        &               &                             \\
\hline
LaCrO$_3$      &          & PC           & 0.1699                                             & 3.75            & 2107\%             & 5                      & $\sim$37.1    & \cite{Fkhar2020}            \\
GdCrO$_3$      & 7        & SC (FZM)     &                                                    & 57.47           &                    & 14                     & 6             & TS                          \\
GdCrO$_3$      & 7        & SC (FLM)     & 29.5                                               & 35.52           & 20.4\%             & 4                      & 3             & \cite{Yin2015-2}            \\
GdCrO$_3$      & 7        & PC           & 41.24                                              & 49.96           & 21.1\%             & 9                      & 3.8           & \cite{Mahana2018-1}         \\
TbCrO$_3$      & 9        & SC (FLM)     & 5.0                                                & 11.97           & 139.4\%            & 2                      & $\sim$4.5     & \cite{Yin2016-2}            \\
DyCrO$_3$      & 10       & PC           & 10.85                                              & 28.70           & 164.5\%            & 4                      & 5             & \cite{McDannald2015}        \\
HoCrO$_3$      & 11       & PC           & 7.2                                                & 18.72           & 160\%              & 7                      & 20            & \cite{Yin2015}              \\
ErCrO$_3$      & 12       & PC           & 10.7                                               & 22.88           & 113.8\%            & 7                      & 15            & \cite{Shi2018}              \\
TmCrO$_3$      & 13       & PC           & 4.6                                                & 16.6            & 260.8\%            & 5                      & $\sim$13.4 K  & \cite{Yoshii2012}           \\
YbCrO$_3$      & 14       & PC           & $\sim$1.91                                         & 18.23           & 854.5\%            & 5                      & $\sim$12.33   & \cite{Oliveira2017}         \\
\end{tabular}
\end{ruledtabular}
\end{table*}

To gain in-depth understanding of the two magnetic transitions occurring at 2.33 and 6.74 K, we further fit the temperature-dependent (from 7 to 30 K) ZFC magnetization data using Eq.~(\ref{realMT}). The $m$ value in Eq.~(\ref{realMT}) is determined by the CW constant. We therefore fixed the parameters of $m$ and $\Theta_{\textrm{CW}}$ (Table~\ref{gamma}) and released only $M_\textrm{BG}$ and $\gamma$ for the fits. To explore the detailed temperature-dependent values of $M_\textrm{BG}$ and $\gamma$, we divided the temperature range of 7--30 K into five regimes, i.e., 7--10, 10--15, 15--20, 20--25, and 25--30 K. The fit results are listed in Table~\ref{gamma}. It is noted that as temperature decreases from 30 to 7 K, the value of $\gamma$ increases, consistent with the hypothesis that there exist short-range AFM orders of Gd$^{3+}$ ions above $T_{\textrm{N-Gd}}$. It is more interesting that the value of $M_\textrm{BG}$ changes from negative (15--30 K) to positive (7--15 K). This sign change most likely indicates that the AFM axis of Cr$^{3+}$ ions rotates, consistent with the observed spin-reorientation transition.

\subsection{C. Magnetic phase diagrams}

Figure~\ref{phasediag} shows the measurements of magnetization as a function of temperature at different applied magnetic fields. The values of $T_{\textrm{N-Gd}}$ and $T_{\textrm{SR}}$ were determined as shown in Fig.~\ref{magT}(c). The indications of $T_{\textrm{N-Gd}}$ and $T_{\textrm{SR}}$ were clearly observed at 300--1000 Oe [Fig.~\ref{phasediag}(a)]. With increasing applied magnetic field, the values of $T_{\textrm{N-Gd}}$ and $T_{\textrm{SR}}$ shift to lower and lower temperatures [Figs.~\ref{phasediag}(b) and \ref{phasediag}(c)]. Temperature-dependent magnetization curves around $T_{\textrm{SR}}$ were previously measured at 0, 50, 500, and 1000 Oe \cite{Cooke1974, Yin2015-2}. Unfortunately, due to the presence of impurities \cite{Cooke1974, Yin2015-2}, the indication of the SR transition disappeared when the strength of applied magnetic field was stronger than 1000 Oe, and the kink indicative of the AFM transition of Gd$^{3+}$ ions at $\sim$2.3 K did not appear \cite{Yin2015-2}. By comparison, our study clearly shows both features with the high-quality GdCrO$_3$ single crystals. This makes us confident to deeply explore the detailed magnetic phase diagram.

As shown in Fig.~\ref{phasediag}(c), above $\sim$0.68 T, the value of $T_{\textrm{N-Gd}}$ combines with that of $T_{\textrm{SR}}$, and the indication of $T_{\textrm{N-Gd}}$ is indistinguishable; above $\sim$0.8 T, the signature of $T_{\textrm{SR}}$ disappears, indicating that there exist applied-magnetic-field-driven magnetic phase transitions \cite{HFLi2016}. We therefore divided the magnetic phase diagram [Fig.~\ref{phasediag}(c)] into four regimes. Within the regime 1, the ions of magnetic Cr$^{3+}$ form a long-range ordered AFM structure with a small canting [Fig.~\ref{magT}(b)]. With decreasing temperature and increasing applied magnetic field, the AFM easy axis $M^0_+M^0_-$ of Cr$^{3+}$ magnetic ions changes from one direction to another (regime 2), depending on the competing degree between anisotropic exchange and single-ion anisotropic energies of Cr$^{3+}$ ions as previously predicated theoretically \cite{HFLi2016}. While further decreasing temperature, the long-range magnetic order of Gd$^{3+}$ ions forms (regime 3). Regime 4 has not been explored yet owing to the technique limitation. The applied magnetic field shifts $T_{\textrm{N-Gd}}$ to lower temperatures, resembling the behavior of a normal antiferromagnet. It is abnormal that the applied magnetic field also suppresses the values of $T_{\textrm{SR}}$, which necessitates a further hot-neutron scattering study to solve this puzzle. It is pointed out that the magnetic phase diagram of applied magnetic field and temperature [Fig.~\ref{phasediag}(c)] was compiled with the measurements of magnetization, which strongly depends on the relative magnetic contributions of Gd$^{3+}$ and Cr$^{3+}$ ions \cite{Yoshii2001}.

As shown in Fig.~\ref{phasediag}(d), at 50 Oe, the $M-T$ curve behaves like an antiferromagnet. Upon cooling, the magnetization increases obviously at $T_{\textrm{N-Cr}} =$ 168.86(2) K, reaches a maximum at 168.40(2) K, and subsequently decreases sharply and then becomes negative at 167.99(2) K, followed by a smooth decrease until $\sim$162 K. After that, the magnetization increases again and becomes positive at $\sim$149.73 K. A Similar magnetic reversal was also observed in TmCrO$_3$ \cite{Yoshii2012, Yoshii2019}, EuCr$_{0.85}$Mn$_{0.15}$O$_3$ \cite{Kumar2017}, and YbCrO$_3$ \cite{Su2010} compounds. By contrast, above 150 Oe, the values of the measured magnetization always remain positive [Figs.~\ref{phasediag}(d) and \ref{phasediag}(e)], and the magnetization in the temperature range from $\sim$140 K to $T_{\textrm{N-Cr}}$ becomes larger and larger with increasing applied magnetic field, so that the kink indicative of the appearance of $T_{\textrm{N-Cr}}$ [Fig.~\ref{magT}(d)] gets weaker and weaker and finally disappears above $\sim$4.3 T [Figs.~\ref{phasediag}(e) and \ref{phasediag}(f)]. The kink may be buried in the higher magnetization signal induced by higher applied magnetic fields, or its disappearance might indicate a meltable magnetic state. Utilizing the method shown in Fig.~\ref{magT}(d), we determined the values of $T_{\textrm{N-Cr}}$ as a function of applied magnetic field, as shown in Fig.~\ref{phasediag}(f). The applied magnetic fields enhance the values of $T_{\textrm{N-Cr}}$, consistent with the hypothesis that below $T_{\textrm{N-Cr}}$ the magnetic Cr$^{3+}$ ions order with a canted AFM structure.

Figure~\ref{magH}(a) shows the ZFC magnetization as a function of applied magnetic field from --14 to 14 T at marked temperatures. At 1.8 K, the magnetization increases almost linearly from 0 to $\sim$2 T and then approaches towards a saturation magnetic state above $\sim$4 T. At 5.1 K, the saturated magnetization at 14 T is $\sim$4\% higher than that at 1.8 K, consistent with the formation of a Gd$^{3+}$ magnetic structure below $T_{\textrm{N-Gd}}$. We transferred the unit of magnetization from emu per gram into $\mu_\textrm{B}$ per chemical formula unit (GdCrO$_3$) and found that the values of the saturation moments under 14 T were $\sim$6.43 $\mu_\textrm{B}$ (at 1.8 K) and $\sim$6.69 $\mu_\textrm{B}$ (at 5.1 K). These values are a little smaller than the theoretical saturation moment of Gd$^{3+}$ ions, i.e., $g_JJ =$ 7 $\mu_\texttt{B}$, consistent with the foregoing discussions that the Gd$^{3+}$ ions in the GdCrO$_3$ compound form a long-range magnetic order. To check a possible magnetic hysteresis effect, we measured the magnetization from --1.2 to 1.2 T in detail as temperature decreased from 180 to 3.2 K [Figs.~\ref{magH}(b)-\ref{magH}(d)]. As shown in Fig.~\ref{magH}(b), no hysteresis loop was observed at 180 K. Upon cooling, it appears at 167 K (below $T_\textrm{N-Cr}$), consistent with the hypothesis that the Cr$^{3+}$ ions in the GdCrO$_3$ compound form a canted AFM structure. Upon further cooling down to 60 K [Fig.~\ref{magH}(c)], the magnetic hysteresis effect gets more and more obvious with enhanced remanent magnetization. As shown in Fig.~\ref{magH}(d), at 8 K, the hysteresis loop becomes very small, and the magnetization shows a nonlinear field dependence, a characteristic feature of ferromagnetism or a short-range AFM state \cite{Li2015, HFLi2016}. At 5.1, 3.2, and 1.8 K (below $T_{\textrm{SR}}$), the magnetic hysteresis loops are indistinguishable. This may indicate that the canting degree of Cr$^{3+}$ magnetic sublattice becomes very small or a magnetic phase transition for Cr$^{3+}$ ions from the canted to a collinear AFM structure exists.

\subsection{D. Magnetocaloric effect}

To study the MCE of our grown GdCrO$_3$ single crystals, we measured field-dependent magnetization from 0 to 14 T at temperatures indicated in Fig.~\ref{magH}(e). The magnetic entropy change --${\Delta}S_{\textrm{M}}$ can be calculated by
\begin{eqnarray}
|{\Delta}S_{\textrm{M}}(T, \mu_0H)| = \mu_0\sum\limits_{i}\frac{M_{i+1}-M_{i}}{T_{i+1}-T_{i}}{\Delta}H_{i},
\label{Mentropy}
\end{eqnarray}
where $|{\Delta}S_{\textrm{M}}|$ is the absolute value of the magnetic entropy change, ${\mu}_{\textrm{0}}$ is the permeability of vacuum, $M_{i+1}$ and $M_{i}$ represent measured values of magnetization at temperatures of $T_{i+1}$ and $T_{i}$, respectively, and ${\Delta}H_{i}$ is the differential element of the applied magnetic field. Based on Eq.~(\ref{Mentropy}), we calculated the values of --${\Delta}S_{\textrm{M}}$ of the single-crystal GdCrO$_3$ compound, and the results are shown in Fig.~\ref{magH}(f). It is clear that below 9 T, the value of --${\Delta}S_{\textrm{M}}$ reaches a maximum at $\sim$4 K; above 9 T, the maximum point shifts to an elevated temperature $\sim$6 K. For example, at 6 K and 14 T, --${\Delta}S_{\textrm{M}}$ $\approx$ 57.47 J/kg K, which decreases rapidly upon warming. With a field change of ${\Delta}{\mu}_0H$ = 9 T, we calculated the magnetic entropy change --${\Delta}S_{\textrm{M}}$ = 49.11 J/kg K at 4 K for the single-crystal GdCrO$_3$ sample. This value is $\sim$19.1\% higher than the value 41.24 J/kg K measured with a polycrystalline GdCrO$_3$ sample under the same conditions at 3.8 K \cite{Mahana2018-1}. We compared the magnetic entropy changes of GdCrO$_3$ single crystals grown with two different methods: One was the flux method, where --${\Delta}S_{\textrm{M}}$ = 29.5 J/kg K with ${\Delta}{\mu}_0H$ = 4 T at 3 K \cite{Yin2015-2}; the second one was the laser-diode-heated FZ technique from the present study where --${\Delta}S_{\textrm{M}}$ ${\approx}$ 35.52 J/kg K (extrapolated) with the same values of ${\Delta}{\mu}_0H$ and temperature, an improvement of approximately 20.4\%. It is obvious that the single-crystal GdCrO$_3$ compound synthesized by the FZ method \cite{Wu2020-1, Zhu2019-1, Zhu2020} shows a much stronger MCE than the polycrystalline samples, and is even much better than the GdCrO$_3$ single crystal grown by the flux method \cite{Yin2015-2}. In Table~\ref{MCE}, we further compare our results with those from other \emph{R}CrO$_3$ compounds. For example, at 20 K and 7 T, we calculate --${\Delta}S_{\textrm{M}}$ $\approx$ 18.72 J/kg K for the GdCrO$_3$ single crystal. This is $\sim$160\% larger than that of the HoCrO$_3$ compound, 7.2 J/kg K \cite{Yin2015}. We also compared our results with those from other studies on DyCrO$_3$ \cite{McDannald2015}, ErCrO$_3$ \cite{Shi2018}, etc., as listed in Table~\ref{MCE}. This demonstrates that the single-crystal GdCrO$_3$ compound with enhanced MCE is a promising material for potential application in magnetic refrigerators.

With our measured temperature-dependent specific-heat data at different applied magnetic fields, as shown in Fig.~\ref{Magcal}(a), the adiabatic temperature change $\Delta$$T_\textrm{ad}$ can be calculated according to \cite{Rostamnejadi2011,Mahana2017}
\begin{eqnarray}
{\Delta}T_{\textrm{ad}} = \int^{\mu_0 H}_0\frac{T}{C_p(T,{\mu_0 H})}\mathlarger{\mathlarger{\mathlarger{(}}}\frac{\partial M}{\partial T}\mathlarger{\mathlarger{\mathlarger{)}}}_{\mu_0 H}d{\mu_0 H}.
\label{DeltaT-ad}
\end{eqnarray}
We calculated the temperature dependence of $\Delta$$T_\textrm{ad}$ as well as the applied magnetic-field-dependent maximum $\Delta$$T^\textrm{max}_\textrm{ad}$, as shown in Figs.~\ref{Magcal}(b) and \ref{Magcal}(c), respectively. Figure~\ref{Magcal}(b) shows an applied-magnetic-field-driven enhancement of $\Delta$$T_\textrm{ad}$ when $\mu_0H \geq$ 10 T. The grown GdCrO$_3$ single crystal in this study holds large values of adiabatic temperature change, e.g., $\Delta$$T_\textrm{ad}^\textrm{max}$ $\approx$ 16.40 K at 8 T and 24.69 K at 14 T. The $\Delta$$T_\textrm{ad}^\textrm{max}$ value of the single-crystal GdCrO$_3$ compound is much larger than that of other Gd-based perovskites such as GdMnO$_3$ and GdAlO$_3$ \cite{Mahana2017}. It is also larger than that of some lanthanide-based oxides such as  EuHo$_2$O$_4$ ($\Delta$$T_\textrm{ad}^\textrm{max}$ $\approx$ 12.7 K) and EuDy$_2$O$_4$ ($\Delta$$T_\textrm{ad}^\textrm{max}$ $\approx$ 16 K) \cite{Midya2012}.

\section{IV. Conclusions}

In summary, we have investigated the structural and magnetic properties of the GdCrO$_3$ single crystal. The collected XRPD pattern was well indexed with the space group $Pmnb$, from which we extracted the lattice constants and atomic positions. The magnetization data at 500 Oe from 200 to 300 K agree well with the CW law, which results in a PM CW temperature $\Theta_{\textrm{CW}} =$ --20.33(4) K and an effective PM moment 8.40(9) $\mu_\textrm{B}$. Taking into account both the temperature-dependent and -independent net-magnetization of Cr$^{3+}$ ions, we can fit well the magnetization data at 30--140 K. We clearly observed the indications of the formation of the canted AFM structure of Cr$^{3+}$ ions at $T_{\textrm{N-Cr}}$, the spin reorientation of Cr$^{3+}$ moments at $T_{\textrm{SR}}$, and the formation of a long-range-ordered Gd$^{3+}$ magnetic structure at $T_{\textrm{N-Gd}}$. We have constructed the magnetic phase diagrams of $T_{\textrm{N-Cr}}$, $T_{\textrm{SR}}$, and $T_{\textrm{N-Gd}}$ as a function of applied magnetic field and proposed magnetic configurations in the corresponding temperature regimes. The magnetic phase diagrams of $T_{\textrm{N-Gd}}$ and $T_{\textrm{N-Cr}}$ are consistent with the corresponding magnetic structures, whereas, the magnetic phase diagram of $T_{\textrm{SR}}$ seems to be abnormal. When $T_{\textrm{SR}} < T < T_{\textrm{N-Cr}}$, obvious magnetic hysteresis loops were observed. Below $T_{\textrm{SR}}$, the hysteresis loop becomes very weak, probably indicating a decrease in the canting degree of the Cr$^{3+}$ magnetic structure or a phase transition from the canted to a collinear AFM structure. We calculated the magnetic entropy change --$\Delta S_\textrm{M}$. For example, --$\Delta S_\textrm{M} \approx$ 35.52 J/kg K at ${\Delta}{\mu}_0H =$ 4 T and 3 K, an improvement of $\sim$20.4\% compared with that of the GdCrO$_3$ single crystal grown by the flux method \cite{Yin2015-2}. We summarized the MCE of \emph{R}CrO$_3$ compounds (\emph{R} = $4f^n$ rare earths, $n =$ 7--14) and found that our grown GdCrO$_3$ single crystal displayed the highest value of magnetic entropy change and an enhanced adiabatic temperature change. This indicates that the single-crystal GdCrO$_3$ compound is a potential candidate for magnetic cooling.

The natural Gd atom is a very strong neutron absorber. Therefore, uniquely determining the interesting magnetic structures, as well as the magnetic phase-transition diagrams explored in this study, necessitates a single-crystal neutron-diffraction study with a hot neutron source.

\textbf{Acknowledgments}

T.L. acknowledges the National Natural Science Foundation of China (Grant No. 11604214), the Foundation of Department of Education of Guangdong Province (Grant No. 2018KTSCX223), and the Foundation of Department of Science and Technology of Guangdong Province (Grant No. 2020A1515010814).
Z.L. and Z.C. acknowledge the Key-Area Research and Development Program of Guangdong Province (2020B090922006) and the Key Project of Natural Science Foundation of China (61935010, 61735005).
H.-F.L. acknowledges the University of Macau (Files Nos. SRG2016{--}00091{--}FST and MYRG2020--00278--IAPME), the Science and Technology Development Fund, Macao SAR (File Nos. 063/2016/A2, 064/2016/A2, 028/2017/A1, and 0051/2019/AFJ), and the Guangdong--Hong Kong--Macao Joint Laboratory for Neutron Scattering Science and Technology.

\end{document}